\documentclass{article}
\usepackage{spconf,amsmath,graphicx}
\usepackage{graphicx}
\usepackage{multirow,booktabs,amssymb}
\def\x{{\mathbf x}}
\def\L{{\cal L}}

\title{Learning from yourself: a Self-distillation method for fake speech detection}
\name{Jun Xue$^{1}$, Cunhang Fan$^{1,*}$\thanks{$^{*}$ Corresponding authors.}, Jiangyan Yi$^{2}$, Chenglong Wang$^{2}$, Zhengqi Wen$^{3}$, Dan Zhang$^{4}$, Zhao Lv$^{1,*}$}

\address{
	$^1$Anhui Province Key Laboratory of Multimodal Cognitive Computation, \\
	School of Computer Science and Technology, Anhui University, Hefei 230601, China \\
	$^2$NLPR, Institute of Automation, 
	Chinese Academy of Sciences, Beijing 100190, China\\
	$^3$Qiyuan Laboratory, Beijing 100190, China\\
	$^4$Department of Psychology, Tsinghua University, Beijing 100190, China\\
	\small{
		e21201068@stu.ahu.edu.cn, \{cunhang.fan, kjlz\}@ahu.edu.cn,  \{jiangyan.yi,chenglong.wang\}@nlpr.ia.ac.cn,}\\ \small{wenzhengqi@qiyuanlab.com, dzhang@tsinghua.edu.cn}
}
\begin{document}
	%
	\maketitle
	\begin{abstract}
In this paper, we propose a novel self-distillation method for fake speech detection (FSD), which can significantly improve the performance of FSD without increasing the model complexity. 
For FSD, some fine-grained information is very important, such as spectrogram defects, mute segments, and so on, which are often perceived by shallow networks.
However, shallow networks have much noise, which can not capture this very well. To address this problem, we propose using the deepest network instruct shallow network for enhancing shallow networks.
Specifically, the networks of FSD are divided into several segments, the deepest network being used as the teacher model, and all shallow networks become multiple student models by adding classifiers.
Meanwhile, the distillation path between the deepest network feature and shallow network features is used to reduce the feature difference. 
A series of experimental results on the ASVspoof 2019 LA and PA datasets show the effectiveness of the proposed method, with significant improvements compared to the baseline.

	\end{abstract}
	\begin{keywords}
		Fake speech detection, self-distillation, automatic speaker
		veriﬁcation, ASVspoof
	\end{keywords}
	\section{Introduction}
	\label{sec:intro}
	
	
	With the rise of biometrics, automatic speaker verification (ASV) has also started to be widely used. However, the development of synthetic speech technology seriously threatens the security of ASV systems. The main attack types for ASV systems are audio replay, text-to-speech (TTS), and voice conversion (VC). Therefore, a series of fake speech detection (FSD) challenges are used to improve the security of ASV systems.
	
	
	\begin{figure*}[h]	
		\centering
		\includegraphics[width=1.0\textwidth]{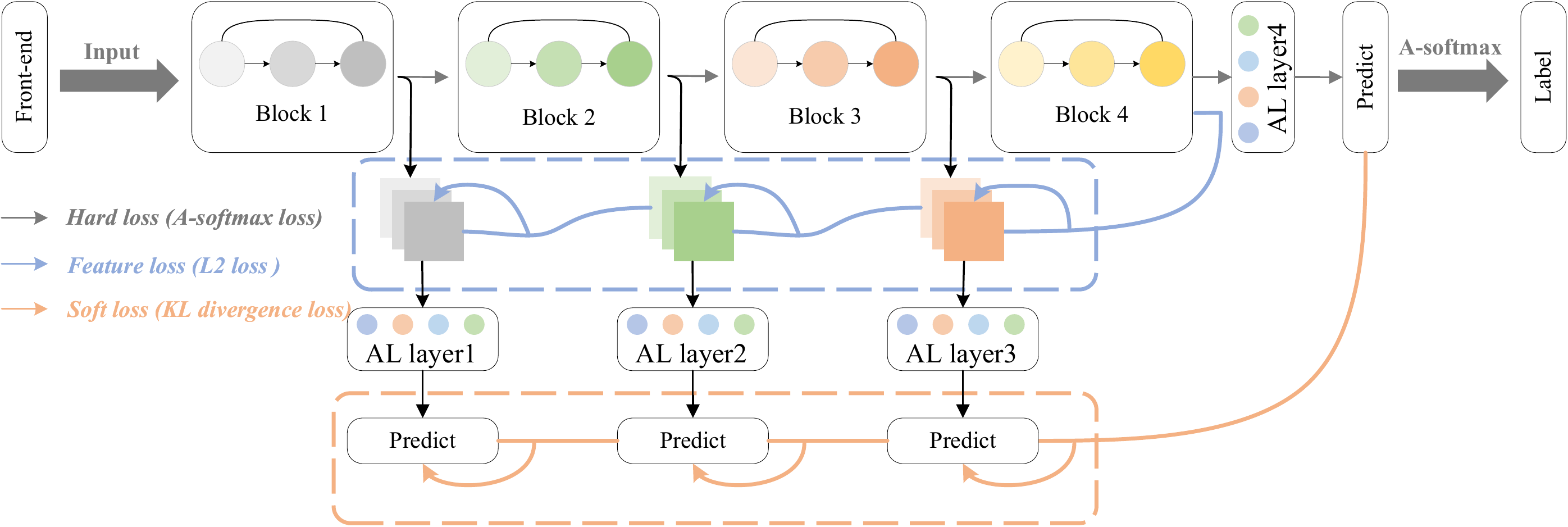}
		
		\caption {The proposed self-distillation framework is based on SENet and ECANet networks. The framework is divided into four blocks and each block is set with a classifier (AngleLinear, AL). Note: The additional classifiers can all be removed during evaluation and thus do not increase the model load.}
		\label{fig:fig2}
	\end{figure*}
	
	The research of FSD revolves around finding discriminately features and designing robust networks. The front-end features mainly are raw waveforms~\cite{jung2019rawnet,tak2021end2} , linear frequency cepstral coefficients (LFCC), log power spectrogram (LPS), and so on. For different front-end features, researchers have proposed a series of convolutional neural network modeling methods.
	Most of the networks are based on ResNet, LCNN \cite{wang2022fully}, and graph networks.
	To further improve the model performance, researchers introduced attention mechanisms into different networks.  Ling et al. \cite{Ling2021} proposed an attention-based convolutional neural network for the FSD task, authors designed a frequency attention block and a channel attention block to focus the discriminative information of the features. In \cite{Tak2021GraphAN}, they first time applied graph attention (GAT) networks to FSD tasks to further analyze the relationship between adjacent subbands in spectral features. Jung et al. \cite{jung2022aasist} proposed a GAT-based architecture using a heterogeneous attention mechanism and stack nodes to integrate temporal and spectral domains.
	Zhang et al. \cite{Zhang} used the squeeze and excitation (SE) network for FSD, which is the SE part for computing global channel attention weights. In addition, efficient channel attention (ECA) \cite{Wangme} is proposed to solve the problem of information loss due to SE block dimensionality reduction, which is widely used in other fields.
	
	In addition, self-distillation as popular methods has gained good performance in many fields. In \cite{xu2022self}, the authors proposed an attention-based feature self-distillation method, which can utilize more high-level information to improve the performance of automatic speech recognition. Liu et al. \cite{liu2022self} designed a feature-enhanced self-distillation method for speaker verification, which achieved good performance. Ge et al. \cite{ge2020bake} proposed an integrated propagation of knowledge form to design a self-distillation framework, which is very effective for image classification tasks.
	
	In this paper, we propose a novel self-distillation approach for FSD. 
Many studies show that the shallow feature of voice is important for FSD. In [7], the authors find that mute segments of voices affect the performance of FSD seriously. 
Further, Deng et al. \cite{deng2022detection} utilize shallow feature information by designing simple classifiers, such as spectrum defects, mute segments, and so on. 
While shallow networks are sensitive to this kind of information, the capture ability is not well as deep networks.
To solve this problem, we propose a self-distillation framework that the deepest network instructs shallow networks, which can further strengthen shallow networks.
Specifically, during training, we add classifiers as multiple student models for all shallow networks and set the deepest network as the teacher model. Based on the prediction results, the teacher model instructs the student model to transform the deepest knowledge into all shallow networks. Meanwhile, to reduce the feature difference between the shallow and deepest networks, we also build self-distillation models in the feature dimension.
It is worth noting that the classifiers in all shallow layers can be removed during inference, thus causing no additional computational and model complexity. To validate the effectiveness of the proposed approach, we used the ECANet and SENet networks as the baseline and the LA and PA from the ASVspoof 2019 challenge as the dataset. Compared to the baseline, the performance of the system is significantly improved after using self-distillation.

	\section{The Proposed self-distillation Method}
	
	\subsection{Self-distillation}
	\label{sec:format}
	In this section, we propose a self-distillation method as shown in Fig.\ref{fig:fig2}. Specifically, Inspired by \cite{zhang2019your}, we divide it into four segments based on the model architecture. In this paper, we divide the model into four blocks, and the block is SE block or ECA block. Then, we set a classifier after each block, and we use AngleLinear to obtain the prediction results of each layer. Note that these extra classifiers are only used in training and do not increase the load during inference. In the training phase, we use the fourth layer of the network as the teacher model and the first three layers as the student model. The deep network knowledge is transformed to the shallow layer in both feature and prediction dimensions.
	
	To make full use of the network information, three losses during the training process are introduced:

	\begin{table*}[t]
		\caption{EER and t-DCF results for different architectures based on the ASVspoof 2019 LA and PA dataset. SD indicates that the self-distillation method is used.}
		\label{tab:result1}
		\resizebox{\linewidth}{!}{
			\begin{tabular}{ccccccccccc}
				\hline
				\multirow{5}{*}{LA dataset} & \multicolumn{2}{c}{Model}         & ECANet9 & ECANet18         & ECANet34         & ECANet50 & SENet9 & SENet18 & SENet34         & SENet50         \\ \cline{2-11} 
				& \multirow{2}{*}{EER}   & baseline & 1.48   & 1.18            & 1.44            & 1.87    & 1.97   & 1.52    & 1.23            & 1.83            \\ \cline{3-11} 
				&                        & SD       & 1.22   & \textbf{0.88}   & 1.14            & 1.09    & 1.23   & 1.37    & \textbf{1.08}   & \textbf{1.00}   \\ \cline{2-11} 
				& \multirow{2}{*}{t-DCF} & baseline & 0.0493 & 0.0378          & 0.0460          & 0.0605  & 0.0610 & 0.0497  & 0.0358          & 0.0536          \\ \cline{3-11} 
				&                        & SD       & 0.0376 & \textbf{0.0295} & 0.0334          & 0.0318  & 0.0388 & 0.0417  & \textbf{0.0347} & \textbf{0.0309} \\ \hline
				\multirow{5}{*}{PA dataset} & \multicolumn{2}{c}{Model}         & ECANet9 & ECANet18         & ECANet34         & ECANet50 & SENet9 & SENet18 & SENet34         & SENet50         \\ \cline{2-11} 
				& \multirow{2}{*}{EER}   & baseline & 0.83   & 0.95            & 0.88            & 0.90    & 0.97   & 0.93    & 1.14            & 0.82            \\ \cline{3-11} 
				&                        & SD       & 0.74   & 0.83            & \textbf{0.70}   & 0.82    & 0.85   & 0.87    & \textbf{0.65}   & 0.79            \\ \cline{2-11} 
				& \multirow{2}{*}{t-DCF} & baseline & 0.0221 & 0.0284          & 0.0255          & 0.0262  & 0.0269 & 0.0266  & 0.0334          & 0.0228          \\ \cline{3-11} 
				&                        & SD       & 0.0199 & 0.0219          & \textbf{0.0208} & 0.0222  & 0.0232 & 0.0239  & \textbf{0.0174} & 0.0219          \\ \hline
			\end{tabular}
		}
	\end{table*}

	\begin{table}[t]
		\caption{The ECANet and SENet model architecture and configuration. Dimensions refer to (channels, frequency, and time). Batch normalization (BN) and Rectified Linear Unit (ReLU). ECA and SE are the efficient channel attention block and the squeeze and excitation block, respectively.}
		\label{model}
		\resizebox{\linewidth}{!}{
			\begin{tabular}{ccc}
				\hline
				Layer                            & Input: 27000 samples & Output shape  \\ \hline
				Front-end                        & F0 subband           & (45,600)(F,T) \\ \hline
				\multirow{3}{*}{Post-processing} & Add channel          & (1,45,600)    \\
				& Conv2D\_1            & (16,45,600)   \\
				& BN \& ReLU           &               \\ \hline
				Block1                       &  $\begin{matrix}
					c_{1} \times\left\{\begin{array}{c}
						
						\text { Conv2D\_3 } \\
						\text { Conv2D\_3 }  \\
						\text {ECA or SE}

					\end{array}\right\}		 
				\end{matrix}$            & (32,45,600)   \\ \hline
				Block2                       &  $\begin{matrix}
					c_{2} \times\left\{\begin{array}{c}
						
						\text { Conv2D\_3 } \\
						\text { Conv2D\_3 }  \\
						\text {ECA or SE}

					\end{array}\right\}		 
				\end{matrix}$            & (64,23,300)   \\ \hline
				Block3                       &  $\begin{matrix}
					c_{3} \times\left\{\begin{array}{c}
						
						\text { Conv2D\_3 } \\
						\text { Conv2D\_3 }  \\
						\text {ECA or SE}

					\end{array}\right\}		 
				\end{matrix}$            & (128,12,150)  \\ \hline
				Block4                       &  $\begin{matrix}
					c_{4} \times\left\{\begin{array}{c}
						
						\text { Conv2D\_3 } \\
						\text { Conv2D\_3 }  \\
						\text {ECA or SE}

					\end{array}\right\}		 
				\end{matrix}$            & (256,6,75)    \\ \hline
				\multirow{2}{*}{Output}          & Avgpool2D(1,1)       & (256,1,1)     \\
				& AngleLinear          & 2             \\ \hline
			\end{tabular}
		}
	\end{table}

	\begin{itemize}
\vspace{-0.33ex}
		\item 
		Hard loss: The A-softmax function is used to calculate the loss of the labels and the fourth layer classifier. This is calculated using the labels of the training dataset and the output from the Anglelinear classifier, which is used to fully extract the hidden knowledge in the training dataset.
\vspace{-0.33ex}
		\item  
		Feature loss: The L2 function is used to compute the feature mapping between the fourth layer network and each shallow layer network. This can introduce non-explicit knowledge of the deepest features into the shallow features so that their shallow networks can match better with the deeper networks when predicting.
\vspace{-0.33ex}
		\item
		Soft loss: The KL divergence function is used to calculate the soft loss in the teacher-student model. The deepest network output is used as the teacher model and several shallow network outputs are used as the student model. The difference between the distributions of the two outputs is calculated, which can guide the shallow network to learn more.
		
	\end{itemize}

		\vspace{-3ex}
	\subsection{Training Methods}
	When training with the self-distillation method, the loss has three components. First, we compute the hard loss of the deepest network and labels:
	\vspace{-1ex}
	\begin{equation}
		\mathcal{L}_{hard }={A}_{-} {softmax}\left(p^n, L\right)
	\end{equation}
	
	where $p^n$ is the deepest output of the network, in this paper, we set the $n$ is 4.
	$L$ is the label of the training data set, and ${A}_{-} {softmax}$ denotes the A-softmax function. $\mathcal{L}_{hard }$ is the hard loss.
	
	Soft loss is used for knowledge distillation of the shallow and deep networks, and we calculate KL divergence using each shallow and deepest network.
	\vspace{-1.2ex}
	\begin{equation}
		\mathcal{L}_{soft }=\sum_i^{n-1} KL\left(p^i, p^n\right)
	\end{equation}
	
	$KL$ denotes the KL divergence function and $p^i$ is the output of each layer of the network after the Angellinear classifier. $\mathcal{L}_{soft }$ is the final soft loss.

	The feature loss is used to balance the difference between the shallow and deepest networks, which can be fed back to the classification output of the shallow network to facilitate the soft loss fit.
		\vspace{-2ex}
	\begin{equation}
		\mathcal{L}_{feature }=\sum_i^{n-1} L 2\left(\mathcal{F}^i, \mathcal{F}^n\right)
	\end{equation}
	
	$\mathcal{F}^i$ is the output feature of each layer, and $\mathcal{F}^n$ is the output feature of the deepest layer. $L 2$ is the L2 loss function, and $\mathcal{L}_{feature }$ is the final feature loss.

		\vspace{-2ex}
	\begin{equation}
		\mathcal{L}=\alpha * \mathcal{L}_{hard }+(1-\alpha) * \mathcal{L}_{soft }+\beta * \mathcal{L}_{feature }
	\end{equation}
	
	The loss at training consists of the following three components, $\alpha$ and $\beta$ are hyperparameters to balance the three sources of loss. $\mathcal{L}$ is the final loss.

	\begin{table*}[t]
		\caption{Comparison of our self-distillation system with other known single systems.}
		\label{tab:result2}
		\resizebox{\linewidth}{!}{
			\begin{tabular}{cccc||cccc}
				\hline
				\multicolumn{4}{c||}{LA dataset}                                             & \multicolumn{4}{c}{PA dataset}                                            \\ \hline
				Systems                   & Front-end    & EER(\%)       & t-DCF           & Systems                   & Front-end  & EER(\%)       & t-DCF           \\ \hline
				AASIST \cite{jung2022aasist}                   & Raw waveform            & 0.83          & 0.0275          & T28  \cite{nautsch2021asvspoof}                     & -          & 0.52          & 0.1470          \\ \hline
				ECANet18(SD) \textbf{Ours}   & F0 subband          & \textbf{0.88} & \textbf{0.0295} & SENet34(SD) \textbf{Ours}  & F0 subband        & \textbf{0.65} & \textbf{0.0174} \\ \hline
				SENet50(SD) \textbf{Ours} & F0 subband          & \textbf{1.00} & \textbf{0.0309} & ECANet34(SD) \textbf{Ours} & F0 subband        & \textbf{0.70} & \textbf{0.0208} \\ \hline
				RawGAT-ST \cite{tak2021end}                 & Raw waveform & 1.06          & 0.0340          & SE-Res2Net50 \cite{li2021replay}             & Spec       & 0.74          & 0.0207          \\ \hline
				SENet34(SD) \textbf{Ours}  & F0 subband          & \textbf{1.08} & \textbf{0.0347} & T10  \cite{nautsch2021asvspoof}                        & \textbf{-} & 1.08          & 0.1598          \\ \hline
				FFT-L-SENet  \cite{Zhang}             & LPS          & 1.14          & 0.0368          & T45 \cite{nautsch2021asvspoof}                      & \textbf{-} & 1.23          & 0.1610          \\ \hline
				MCG-Res2Net50 \cite{li2021channel}            & CQT          & 1.78          & 0.0520          & T44   \cite{nautsch2021asvspoof}                    & \textbf{-} & 1.29          & 0.1666          \\ \hline
				Resnet18-OC-softmax \cite{zhang2021one}      & LFCC         & 2.19          & 0.0590          & T53 \cite{nautsch2021asvspoof}                      & -          & 1.66          & 0.1729          \\ \hline
				ResNet18-GAT-T \cite{Tak2021GraphAN}           & LFB          & 4.71          & 0.0894          & Capsule \cite{luo2021capsule}                  & LFCC       & 2.76          & 0.0730          \\ \hline
			\end{tabular}
		}
	\end{table*}

	\section{Experiments and Results}
	\subsection{Datasets}
	We trained and evaluated models on the ASVspoof 2019 LA and PA datasets. The LA set includes three types of spoofing attacks (TTS, VS, and audio replay), which are divided into 19 attack algorithms (A01-A19). The PA set includes only replay attacks, and there are 27 replay attacks in different acoustic environments. 
	In this paper, EER and the minimum normalized tandem detection cost function (min t-DCF) are used as evaluation metrics for assessing the performance of different systems. 

	\subsection{Training setup and baseline}
	Front-end features: Inspired by \cite{xue2022audio}, we use the F0 subband as our input features. Firstly, we extract the full frequency band of LPS and use the window function as Blackman's Short Time Fourier Transform (STFT), setting the window length and hop length as 1728 and 130 respectively. we fix the frame number as 600 and get the LPS to feature 865×600. Finally, we take the first 45 dimensions of the frequency and finally get the front-end F0 subband feature size of 45×600.
	
	Back-end classifier: As shown in Table \ref{model}, we use SENet and ECANet as deep neural network classifiers. 
	where the $c_{1}-c_{4} $ vectors corresponding to the 9, 18, 34 layers are $(1,1,1,1)$, $(2,2,2,2)$, and $(3,4,6,3)$, respectively.
	The 50-layer network is set to three convolutions in the block, with convolution kernels of 1,3,1. The rest is the same as the 34-layer network.
	For training, we use Adam as the optimizer with parameters $\beta_{1}=0.9$, $\beta_{2}=0.98$, $\epsilon=10^{-9}$ and weight decay $10^{-4}$. The number of the epoch is 32. The two hyperparameters $\alpha$ and $\beta$ are set to 0.7 and 0.3, respectively.

	\subsection{Experiment results on LA dataset}

	Table~\ref{tab:result1} shows the EER and t-DCF of the baseline and self-distillation systems for the ASVspoof 2019 LA dataset. The ``SD'' denotes self-distillation.
	According to Table~\ref{tab:result1}, it can be seen that our self-distillation method significantly outperforms the baseline system. In addition, we can observe several interesting phenomena. First, the baseline system ``ECANet18" has the best performance with an EER of 1.18\%. Even so, the self-distillation method can improve it by 25\%. Second, the performance of the different network architectures decreases significantly as the network gets deeper. For example, the EER of the baseline ``SENet50" is 1.83\%, and the EER of its self-distillation is 1.00\%, which is a 45\% improvement.
	Self-distillation effectively reduces the performance degradation of the FSD due to depth and makes its performance smoother for different network depths.
	Table~\ref{tab:result2} shows the performance of the most recently known SOTA single system, and our best system is ranked second.
	In general, the self-distillation method has the effect of fully exploiting the information of different levels of the network. 
	Further, this method has strong generality and improves for different architectures at different depths.

	\subsection{Experiment results on PA dataset}
	
	Table~\ref{tab:result1} shows the EER and t-DCF of the baseline and the self-distillation system for the ASVspoof 2019 PA dataset. according to Table~\ref{tab:result1}, it can be seen that the self-distillation system outperforms the baseline. The EER of the baseline ``SENet34" is 1.14\%, which is the worst performance in the overall baseline. However, the ``SENet34(SD)" system could obtain an EER of 0.65\%, making it the best-performing system. This may be because the shallow network contains more unexplored information, and thus the self-distillation system has a teacher for guidance, which allows shallow networks to mine more fine-grained information as well.
	In addition, Table~\ref{tab:result2} compares the Top systems on the PA dataset, and our method can also get the second one. This indicates that the self-distillation method is very effective.
	Further, our method can also be adapted to different datasets.

	\section{Conclusions}
In this paper, we propose a novel self-distillation method for FSD tasks. This can further improve the performance of FSD without increasing the load and has generality for networks of different architectures. Specifically, we add classifiers behind the shallow network to build interaction with the deepest network in both feature and prediction dimensions, which enhances shallow networks' ability that captures detailed discriminately information.
The feature distillation aims to reduce the difference between deep and shallow features, and the distillation of the prediction dimension is to fully exploit the information in each layer of the network to further optimize the network.
We use different architectures of ECANet and SENet, and our experimental results on ASVspoof 2019 LA and PA datasets validate the effectiveness and generality of our approach, significantly improving the performance of the baseline.
In the future, we will work on building more low-parameter and highly robust FSD systems.
	\section{Acknowledgements}
This work is supported by the National Key Research and Development Plan of China (No.2020AAA0140003), the National Natural Science Foundation of China (NSFC) (No.62201002, No.61972437), Excellent Youth Foundation of Anhui Scientific Committee (No. 2208085J05), Special Fund for Key Program of Science and Technology of Anhui Province (No. 202203a07020008), the Open Research Projects of Zhejiang Lab (NO. 2021KH0AB06) and the Open Projects Program of National Laboratory of Pattern Recognition (NO. 202200014).
	
	%
	%
	
	\bibliographystyle{IEEEbib}
	\bibliography{strings,refs}

\begin{thebibliography}{10}

\bibitem{jung2019rawnet}
Jee-weon Jung, Hee-Soo Heo, Ju-ho Kim, Hye-jin Shim, and Ha-Jin Yu,
\newblock ``Rawnet: Advanced end-to-end deep neural network using raw waveforms
  for text-independent speaker verification,''
\newblock {\em Proc. Interspeech 2019}, pp. 1268--1272, 2019.

\bibitem{tak2021end2}
Hemlata Tak, Jose Patino, Massimiliano Todisco, Andreas Nautsch, Nicholas
  Evans, and Anthony Larcher,
\newblock ``End-to-end anti-spoofing with rawnet2,''
\newblock in {\em ICASSP 2021}. IEEE, 2021, pp. 6369--6373.

\bibitem{wang2022fully}
Chenglong Wang, Jiangyan Yi, Jianhua Tao, Haiyang Sun, Xun Chen, Zhengkun Tian,
  Haoxin Ma, Cunhang Fan, and Ruibo Fu,
\newblock ``Fully automated end-to-end fake audio detection,''
\newblock in {\em DDAM 2022}, 2022, pp. 27--33.

\bibitem{Ling2021}
Hefei Ling, Leichao Huang, Junrui Huang, Baiyan Zhang, and Ping Li,
\newblock ``{Attention-Based Convolutional Neural Network for ASV Spoofing
  Detection},''
\newblock in {\em Proc. Interspeech}, 2021, pp. 4289--4293.

\bibitem{Tak2021GraphAN}
Hemlata Tak, Jee-weon Jung, Jose Patino, Madhu Kamble, Massimiliano Todisco,
  and Nicholas Evans,
\newblock ``End-to-end spectro-temporal graph attention networks for speaker
  verification anti-spoofing and speech deepfake detection,''
\newblock in {\em Proc. ASVSpoof Challenge}, 2021, pp. 1--8.

\bibitem{jung2022aasist}
Jee-weon Jung, Hee-Soo Heo, Hemlata Tak, Hye-jin Shim, Joon~Son Chung, Bong-Jin
  Lee, Ha-Jin Yu, and Nicholas Evans,
\newblock ``Aasist: Audio anti-spoofing using integrated spectro-temporal graph
  attention networks,''
\newblock in {\em ICASSP 2022}. IEEE, 2022, pp. 6367--6371.

\bibitem{Zhang}
Yuxiang Zhang, Wenchao Wang, and Pengyuan Zhang,
\newblock ``{The Effect of Silence and Dual-Band Fusion in Anti-Spoofing
  System},''
\newblock in {\em Proc. Interspeech}, 2021, pp. 4279--4283.

\bibitem{Wangme}
Qilong Wang, Banggu Wu, Pengfei Zhu, P.~Li, Wangmeng Zuo, and Qinghua Hu,
\newblock ``Eca-net: Efficient channel attention for deep convolutional neural
  networks,''
\newblock {\em CVPR 2020}, pp. 11531--11539, 2020.

\bibitem{xu2022self}
Qiang Xu, Tongtong Song, Longbiao Wang, Hao Shi, Yuqin Lin, Yongjie Lv, Meng
  Ge, Qiang Yu, and Jianwu Dang,
\newblock ``Self-distillation based on high-level information supervision for
  compressing end-to-end asr model,''
\newblock {\em Proc. Interspeech 2022}, pp. 1716--1720, 2022.

\bibitem{liu2022self}
Bei Liu, Haoyu Wang, Zhengyang Chen, Shuai Wang, and Yanmin Qian,
\newblock ``Self-knowledge distillation via feature enhancement for speaker
  verification,''
\newblock in {\em ICASSP 2022}. IEEE, 2022, pp. 7542--7546.

\bibitem{ge2020bake}
Yixiao Ge, Xiao Zhang, Ching~Lam Choi, Ka~Chun Cheung, Peipei Zhao, Feng Zhu,
  Xiaogang Wang, Rui Zhao, and Hongsheng Li,
\newblock ``Self-distillation with batch knowledge ensembling improves imagenet
  classification,''
\newblock {\em arXiv}, 2021.

\bibitem{deng2022detection}
JiaCheng Deng, Terui Mao, Diqun Yan, Li~Dong, and Mingyu Dong,
\newblock ``Detection of synthetic speech based on spectrum defects,''
\newblock in {\em DDAM 2022}, 2022, pp. 3--8.

\bibitem{zhang2019your}
Linfeng Zhang, Jiebo Song, Anni Gao, Jingwei Chen, Chenglong Bao, and Kaisheng
  Ma,
\newblock ``Be your own teacher: Improve the performance of convolutional
  neural net via self distillation,''
\newblock in {\em ICCV 2019}. IEEE Computer Society, 2019, pp. 3712--3721.

\bibitem{nautsch2021asvspoof}
Andreas Nautsch, Xin Wang, Nicholas Evans, Tomi~H Kinnunen, Ville Vestman,
  Massimiliano Todisco, H{\'e}ctor Delgado, Md~Sahidullah, Junichi Yamagishi,
  and Kong~Aik Lee,
\newblock ``Asvspoof 2019: spoofing countermeasures for the detection of
  synthesized, converted and replayed speech,''
\newblock {\em IEEE Transactions on Biometrics, Behavior, and Identity
  Science}, vol. 3, no. 2, pp. 252--265, 2021.

\bibitem{tak2021end}
Hemlata Tak, Jee-Weon Jung, Jose Patino, Madhu Kamble, Massimiliano Todisco,
  and Nicholas Evans,
\newblock ``End-to-end spectro-temporal graph attention networks for speaker
  verification anti-spoofing and speech deepfake detection,''
\newblock in {\em ASVSPOOF 2021, Automatic Speaker Verification and Spoofing
  Countermeasures Challenge}. ISCA, 2021, pp. 1--8.

\bibitem{li2021replay}
Xu~Li, Na~Li, Chao Weng, Xunying Liu, Dan Su, Dong Yu, and Helen Meng,
\newblock ``Replay and synthetic speech detection with res2net architecture,''
\newblock in {\em ICASSP 2021}. IEEE, 2021, pp. 6354--6358.

\bibitem{li2021channel}
Xu~Li, Xixin Wu, Hui Lu, Xunying Liu, and Helen Meng,
\newblock ``{Channel-Wise Gated Res2Net: Towards Robust Detection of Synthetic
  Speech Attacks},''
\newblock in {\em Proc. Interspeech 2021}, 2021, pp. 4314--4318.

\bibitem{zhang2021one}
You Zhang, Fei Jiang, and Zhiyao Duan,
\newblock ``One-class learning towards synthetic voice spoofing detection,''
\newblock {\em IEEE Signal Processing Letters}, vol. 28, pp. 937--941, 2021.

\bibitem{luo2021capsule}
Anwei Luo, Enlei Li, Yongliang Liu, Xiangui Kang, and Z~Jane Wang,
\newblock ``A capsule network based approach for detection of audio spoofing
  attacks,''
\newblock in {\em ICASSP 2021}. IEEE, 2021, pp. 6359--6363.

\bibitem{xue2022audio}
Jun Xue, Cunhang Fan, Zhao Lv, Jianhua Tao, Jiangyan Yi, Chengshi Zheng,
  Zhengqi Wen, Minmin Yuan, and Shegang Shao,
\newblock ``Audio deepfake detection based on a combination of f0 information
  and real plus imaginary spectrogram features,''
\newblock in {\em DDAM 2022}, 2022, pp. 19--26.

\end{thebibliography}
	
\end{document}